\documentclass[twocolumn,twocolappendix]{aastex631}
\usepackage{amsmath,textcomp,gensymb,amsbsy}
\usepackage{xcolor}

\newcommand{\logg}{\mbox{$\log g$}}

\shortauthors{Behmard et al.}
\graphicspath{{./}{figures/}}
\begin{document}

\title{M Dwarf Singles and Multis Have Dissimilar Metallicities \\ Indicating Different Origins}

\author[0000-0003-0012-9093]{Aida Behmard}
\altaffiliation{Flatiron Research Fellow}
\affiliation{Center for Computational Astrophysics, Flatiron Institute, 162 Fifth Ave, New York, NY 10010, USA}
\affiliation{American Museum of Natural History, 200 Central Park West, Manhattan, NY 10024, USA}

\author[0000-0002-3022-6858]{Sheila Sagear}
\affiliation{Center for Computational Astrophysics, Flatiron Institute, 162 Fifth Ave, New York, NY 10010, USA}

\author[0000-0003-4769-3273]{Yuxi(Lucy) Lu}
\affiliation{Department of Astronomy, The Ohio State University, Columbus, 140 W 18th Ave., OH 43210, USA}
\affiliation{Center for Cosmology and Astroparticle Physics, The Ohio State University, 191 W. Woodruff Ave., Columbus, OH 43210, USA}

\author[0000-0003-1445-9923]{Romy Rodríguez Martínez}
\affiliation{Center for Astrophysics $\vert$ Harvard \& Smithsonian, 60 Garden Street, Cambridge, MA 02138, USA}

\author[0000-0003-0872-7098]{Adrian M. Price-Whelan}
\affiliation{Center for Computational Astrophysics, Flatiron Institute, 162 Fifth Ave, New York, NY 10010, USA}

\author[0000-0002-0842-863X]{Soichiro Hattori}
\affiliation{Department of Astronomy, Columbia University, 538 West 120th Street, Pupin Hall, New York, NY 10027, USA}
\affiliation{American Museum of Natural History, 200 Central Park West, Manhattan, NY 10024, USA}

\begin{abstract}
It is unclear if host star metallicities differ between systems with one detected transiting planet (``singles") or multiple detected transiting planets (``multis") around FGK dwarfs. Here, we find strong evidence that they do for M dwarfs. We use a homogeneous sample of FGK and M dwarf chemical abundances to demonstrate that M dwarf multis are significantly more metal-poor than singles ($p$ = 0.0008), even when we only consider small ($<$4 $R_{\oplus}$) planets ($p$ = 0.001). This result emerges at the late-K to M dwarf transition. We observe that M dwarf multis become more metal-poor as a function of increasing planet multiplicity. Additionally, planets in singles have significantly higher eccentricities and shorter orbital periods compared to planets in multis. We interpret these findings as evidence that M dwarf singles lead relatively active dynamical lives, and the different planetary architectures of singles and multis arise from distinct dynamics influenced by metallicity. This may not occur in FGK systems because their more massive protoplanetary disks may supply enough planet-forming material to instigate dynamics that lead to only one transiting planet, even when the host star metallicity is low. Another possibility is that multis are less compact around FGK dwarfs because they form from larger disks, which may often cause them to be observed as singles in transit data.

%We speculate that singles and multis around FGK dwarfs lack an observable metallicity difference because of their larger and more massive protoplanetary disks. For example, more massive disks may ensure that the inventory of planet-forming material is large enough to instigate dynamical evolution that leaves one transiting planet behind, even in the low-metallicity regime. Larger disks with more vertical extent may lead

%M dwarfs with higher metallicities may host protoplanetary disks with more solid material that facilitates giant planet formation. These giants may dynamically disrupt other planets in the system, leaving behind system architectures with only one observable planet. This may not occur in FGK systems because the larger host star mass may correspond to a larger protoplanetary disk with enough solids to form giant planets, even when the host star metallicity is low.
\end{abstract}

%\keywords{stars: abundances}


\section{Introduction} \label{sec:intro}
Our growing exoplanet census has revealed two distinct architecture types in the region near the host star: systems with one known planet, or multiple planets. The NASA \emph{Kepler} mission provided the first large influx of multi-planet discoveries (e.g., \citealt{fabrycky2014}). These \emph{Kepler} multi-planet systems are likely coplanar because they all transit their host star (``multis"). This is backed by planet population models which deduce that mutual inclinations of $<$3$^{\circ}$ between planets in multis are consistent with the \emph{Kepler} yield (e.g., \citealt{lissauer2011}). However, the best-fitting models underpredict the number of systems with single transiting planet detections (``singles") by a factor of three \citep{lissauer2011,hansen2013}. This apparent excess of single-planet systems in the \emph{Kepler} census is often referred to as the ``\emph{Kepler} dichotomy". Its underlying cause is not understood, and raises the question of whether singles and multis stem from the same underlying population. That is, do singles belong to the tail of one underlying distribution that includes multis, or do they belong to a separate population with different formation conditions and/or dynamical histories?

%may be primordial (e.g., protoplanetary disks that produce fewer planets) or dynamical (e.g., planet-planet scattering disrupts initially coplanar orbits) in nature. 

To investigate the source of the \emph{Kepler} dichotomy, many studies have examined potential differences in planet and host star properties between singles and multis. Most have been restricted to systems around FGK dwarfs. Several studies report differences in orbital architectures that suggest singles experience more dynamically volatile histories. For example, \citet{morton2014} found that singles host planets with higher obliquities compared to planets in multis. \citet{limbach2015} examined the orbital eccentricities of planets detected with radial velocity (RV) measurements, and found that eccentricities decrease with planet multiplicity. Similarly, \citet{xie2016} and \citet{van_eylen2019} report that singles have significantly higher eccentricities compared to multis. \citet{latham2011} found differences in planet size, with higher occurrence of small sub-Neptunes in multis ($86^{+2}_{-5}\%$) as opposed to singles ($69^{+2}_{-3}\%$). They interpreted this as evidence that singles more commonly host large planets that disrupt or inhibit formation of coplanar planetary architectures. This is in accordance with \citet{brewer2018b} who found that compact multis are more common around metal-poor hosts (which are less likely to host giant planets as expected from the planet-metallicity correlation). However, \citet{Romero2018} report no significant differences in host star metallicity between singles and multis around FGK dwarfs. Larger samples analyzed by \citet{weiss2018} also do not exhibit differences host star metallicity, or any other host star properties tested ($M_{*}$ or $v \sin i$). These results were interpreted by \citet{weiss2018} as evidence that singles and multis around FGK dwarfs belong to the same underlying population. 

%There is evidence that this may differ for singles and multis around M dwarfs. 
Like FGK dwarfs, M dwarfs also host an excess of singles that point to the \emph{Kepler} dichotomy \citep{ballard2016}. Understanding its source in M dwarf systems is of particular interest; M dwarfs are the most common type of star in the Galaxy (composing $\sim$75\% of stars in the solar neighborhood; \citealt{henry2006}), have a higher small planet occurrence rate compared to FGK dwarfs (e.g., \citealt{howard2012,mulders2015b,ment2023}), and typically host compact ($P$ $<$ 20 days) multi-planet systems of small worlds (e.g, \citealt{dressing2015b,hardegree_ullman2019}). A few studies report evidence of chemical differences between M dwarfs that host singles versus multis. \citet{anderson2021} used parallaxes and photometry to derive metallicity proxies for K- and M-dwarf planet hosts, and found tentative evidence ($p$ = 0.015) that compact multis are more metal-poor than singles.
%This works because distance from the main sequence of a colorâmagnitude diagram correlates with stellar metallicity (e.g., \citealt{mann2019}).
Following that, \citet{rodriguez_martinez2023} was the first study to uncover strong evidence that M dwarf multis are relatively metal-poor using a sample of heterogeneous spectroscopic metallicities ($p$ = 0.001). \citet{wanderley2025} report the same result with a small ($<$50 systems) sample of multis and singles around M dwarfs with homogeneous metallicities ($p$ = 0.002). This suggests that differences in M dwarf chemistry lead to the distinct planetary architectures that characterize singles versus multis. 

To definitively demonstrate that M dwarf singles and multis differ in host star chemistry, we require a large and homogeneous sample of M dwarf metallicities. Here, we use a data-driven method ($Lux$, \citealt{horta2025}) to derive chemical abundances for M dwarfs with Milky Way Mapper (MWM) data from the current phase of the Sloan Digital Sky Survey (SDSS-V; \citealt{kollmeier2017}). We describe our planet sample in Section \ref{sec:planet_sample}, and how we derived chemical abundances for their M dwarf hosts in Section \ref{sec:mdwarf_abundances}. In Section \ref{sec:metallicity_comparison}, we examine the metallicities of our M dwarf singles and multis, and compare to FGK systems. We investigate if the metallicity difference is reflected in other host star or planet properties in Sections \ref{sec:additional_star_properties} and \ref{sec:planet_properties}, respectively. Finally, we discuss these findings and their implications for distinct dynamical pathways in singles and multis in Section \ref{sec:discussion}.

\section{The Planet Sample} \label{sec:planet_sample}

We select a sample of transiting single- and multi-planet systems around cool ($<$4500 K) stars from the NASA Exoplanet Archive \footnote{\url{https://exoplanetarchive.ipac.caltech.edu/}, queried 5/21/2026} \citep{christiansen2025}. We focus on $<$4500 K stars because stellar models struggle to provide reliable chemical abundances within this temperature regime (e.g., \citealt{brewer2016}), making it a relatively unexplored parameter space to investigate with our data-driven abundance model. Thus, we include both M and late-K dwarfs in our initial sample. To only retain M and late-K dwarf systems, we apply type-color sequence relations from \citet{pecaut2013ApJS}:

\begin{equation}\label{eq:equation8}
\begin{split}
G - RP > 0.7 \\
G + 5\textrm{log}(\varpi/100) > 6.53
\end{split}
\end{equation}

\noindent where $G$ and $RP$ are photometric passbands and $\varpi$ is the parallax, all from \emph{Gaia} DR3 \citep{gaiadr3}. We then apply a $<$4500 K cut on host star temperatures from the Exoplanet Archive as well to remove any mid-K dwarfs from our sample. We do this because while the \citet{pecaut2013ApJS} criteria enable us to make an initial homogeneous cut on stellar type, the Exoplanet Archive temperatures, while heterogeneous, are likely more accurate for each single system because most are determined spectroscopically. 
%We also check the Exoplanet Archive log $g$ values to ensure that all host stars are in the dwarf regime ($<$ 4 dex). SHOULD I DO THIS? MANY NAN LOG G 
This results in 370 singles, and 248 planets within 97 multis around M and late-K dwarfs. 

Next, we define a smaller sample that includes observations from the Apache Point Observatory Galactic Evolution Experiment (APOGEE; \citealt{majewski2017,wilson2019,almeida2023}) spectrographs. The APOGEE survey is part of SDSS-V/MWM, which we describe further in Section \ref{sec:sdss_mwm_data}. These observations are needed for inferring late-K and M dwarf chemical abundances with our data-driven framework. We cross-match our full planet sample with the latest SDSS-V data release (DR19), but do not impose the $<$4500 K cut on host star temperatures from the Exoplanet Archive. Instead, we apply a $<$4500 K cut using stellar temperatures from the APOGEE Stellar Parameter and Chemical Abundances Pipeline (ASPCAP; \citealt{garcia2016,dr19}), which are homogeneously derived from APOGEE spectra and considered reliable in the late-K and M dwarf regime \citep{souto2022}. We also impose a signal-to-noise ratio (SNR) cut of $>$50/pix. This leaves us with 137 singles, and 125 planets within 48 multis. Throughout this study, we predominantly work with systems only hosted by M dwarfs, which comprise 90 singles and 83 planets within 32 multis.

%a larger sample without SDSS-V/MWM data that we will use to investigate planet and host star properties other than metallicity. 

\section{Host Star Chemical Abundances} \label{sec:mdwarf_abundances}
To robustly investigate differences in chemistry for late-K and M dwarf hosts of singles versus multis, we require a large sample of homogeneously derived stellar chemical abundances. This is challenging because stars with $<$4500 K temperatures are cool enough to harbor molecules in their atmospheres, which create dense clusters of molecular lines in the optical and near-infrared regions of spectra that stellar models struggle to reproduce (e.g., \citealt{allard1997,rojas_ayala2012,mann2013b}). Several studies have modified spectral line lists and atmosphere fitting methodologies to more accurately capture late-K and M dwarf chemistry (e.g., \citealt{souto2022,hejazi2024,melo2024,wanderley2025}). However, these customized modeling routines are computationally expensive and have only been used to characterize small stellar samples ($\sim$50 stars; \citealt{wanderley2025}) with high-resolution spectra to date. 

Alternatively, we can employ data-driven approaches that do not rely on physical stellar models. Data-driven methods are fast compared to traditional stellar atmosphere modeling, and are thus well-suited to characterizing large stellar samples. One notable example is \emph{The Cannon} \citep{ness2015,casey2016}, which uses a training set of stellar spectra with well-determined ``labels" (e.g., stellar parameters, chemical abundances) to construct a predictive model of the flux at every pixel in the wavelength range of the spectra. \emph{The Cannon} has been successfully used to characterize M dwarf spectra and infer a wide range of chemical abundances \citep{behmard2019,birky2020,galgano2020,rains2024,behmard2023}. Here, we employ a similar model (\emph{Lux}; \citealt{horta2025}) that infers labels from stellar spectra via latent representations (making it more flexible compared to \emph{The Cannon}). \emph{Lux} improves upon limitations in previous data-driven/machine learning frameworks (e.g., \citealt{ness2015,ting2019,amdrae2023,guiglion2024,li2024}); it can account for uncertainties in the training set labels, and train models using partially-labeled data (e.g., stars that have $T_{\textrm{eff}}$ information but no [Fe/H]).

\subsection{Lux Model}
\emph{Lux} is a generative model of stellar spectra and labels. In its fiducial implementation, \emph{Lux} uses linear transformations to compute model-predicted label and spectral flux values, where the labels are generated as

\begin{equation}
\boldsymbol\ell_{n} = \textbf{\emph{A}} \hspace{0.5mm}\boldsymbol{z}_{n} + \textrm{noise}
\end{equation}

\noindent where $\boldsymbol\ell_{n}$ is a vector containing the labels, $\boldsymbol{z}_{n}$ is a vector containing the latent parameters, and \textbf{\emph{A}} is a matrix that projects the latent vector onto the stellar labels for the $n$th star. Similarly, the spectral flux values are generated as 

\begin{equation}
\boldsymbol{f}_{n} = \textbf{\emph{B}} \hspace{0.5mm} \boldsymbol{z}_{n} + \textrm{noise}
\end{equation}

\noindent where $\textbf{\emph{f}}_{n}$ represents the flux vector for the $n$th star. We assume the noise to be Gaussian with known variances. This allows us to express the likelihood of the stellar labels as

\begin{equation}
p(\boldsymbol\ell_{n} | \textbf{\emph{A}}, \boldsymbol{z}_{n}) = \mathcal{N}(\boldsymbol\ell_{n} | \textbf{\emph{A}}\hspace{1mm}\boldsymbol{z}_{n},\sigma^{2}_{\ell,n})
\end{equation}

\noindent where $\sigma_{\ell,n}$ represents the uncertainties on labels for each $n$th star, which are in this case the reported uncertainties on stellar parameters and abundances from ASPCAP. The likelihood for spectral fluxes can similarly be represented as

\begin{equation}
p(\boldsymbol{f}_{n} | \textbf{\emph{B}}, \boldsymbol{z}_{n}, \boldsymbol{s}_{f}) = \mathcal{N}(\boldsymbol{f}_{n} | \textbf{\emph{B}}\hspace{0.5mm}\boldsymbol{z}_{n},\sigma^{2}_{f,n} + \boldsymbol{s}_{f}^{2})
\end{equation}

\noindent where $\sigma_{f,n}$ represents the per-pixel flux uncertainties of the stellar spectra, and $\boldsymbol{s}_{f}$ is a free parameter that represents additional per-pixel scatter in the model, and is intended to capture intrinsic scatter and any unaccounted-for systemic errors in the spectra (e.g., sky lines). The joint likelihood of the stellar labels and flux for a given star $n$ can then be represented as

\begin{equation}
\begin{split}
&p(\boldsymbol\ell_{n},\boldsymbol{f}_{n} | \textbf{\emph{A}}, \textbf{\emph{B}}, \boldsymbol{z}_{n}, \boldsymbol{s}_{f})\\
&\hspace{10mm} =  p(\boldsymbol\ell_{n} | \textbf{\emph{A}}, \boldsymbol{z}_{n}) \hspace{1mm}p(\boldsymbol{f}_{n} | \textbf{\emph{B}}, \boldsymbol{z}_{n}, \boldsymbol{s}_{f}),
\end{split}
\end{equation}

\noindent which can be rearranged and expressed for a set of $N$ stars as the product of per-star likelihoods as follows:

\begin{equation}
\begin{split}
\mathcal{L}(\textbf{\emph{A}}, \textbf{\emph{B}},\{\boldsymbol{z}_{n}\}_{N},\boldsymbol{s}_{f}) &= p(\{\boldsymbol\ell_{n}\}_{N}, \{\boldsymbol{f}_{n}\}_{N} | \textbf{\emph{A}}, \textbf{\emph{B}}, \{\boldsymbol{z}_{n}\}_{N}, \boldsymbol{s}_{f}) \\ 
&= \prod^{N}_{n} p(\boldsymbol\ell_{n},\boldsymbol{f}_{n} | \textbf{\emph{A}}, \textbf{\emph{B}}, \boldsymbol{z}_{n}, \boldsymbol{s}_{f}).
\end{split}
\end{equation}

We optimize this likelihood expression against a training set of spectra and labels. In this training step, we infer the parameters \textbf{\emph{A}}, \textbf{\emph{B}}, and $\boldsymbol{s}_{f}$, as well as the latent vector $\boldsymbol{z}_{n}$. We then use the resulting model with defined parameters to infer labels given a test set of stellar spectra.

\subsection{SDSS-V/MWM Data} \label{sec:sdss_mwm_data}
To construct our \emph{Lux} model for late-K and M dwarf spectra, we use observations from the SDSS-V/MWM survey. This survey includes the APOGEE spectrographs, which have collected over one million high resolution ($R$ $\sim$ 22,500) spectra in the $H$-band (1.51$-$1.7 $\mu$m) to date \citep{majewski2017,wilson2019,almeida2023}. The APOGEE survey is designed to target bright objects and focuses mostly on red giants, but also includes $\sim$50,000 M dwarfs. The current SDSS-V data release (DR19) includes products from the APOGEE Stellar Parameter and Chemical Abundances Pipeline (ASPCAP), namely stellar parameters and abundances for a wide set of elements at typical precision of $<$0.1 dex \citep{garcia2016}. However, ASPCAP relies upon stellar atmosphere models, which makes ASPCAP chemical abundances unreliable for stars with $<$4500 K temperatures, such as late-K and M dwarfs. This motivates our data-driven approach to modeling M dwarf spectra with \emph{Lux} \citep{horta2025}.

\emph{Lux} can be executed on either flux- or continuum-normalized spectra. We carry out continuum normalization on the APOGEE late-K and M dwarf spectra via error-weighted broad Gaussian smoothing with 

\begin{equation} \label{eq:equation5}
\begin{split}
\bar{f}(\lambda_{0}) = \frac{\sum_{j} (f_{j} \sigma_{j}^{-2}w_{j}(\lambda_{0}))}{\sum_{j} (\sigma_{j}^{-2} w_{j}(\lambda_{0}))} \hspace{0.5mm},
\end{split}
\end{equation}

\noindent where $f_{j}$ is the flux at pixel $j$ of the wavelength range, $\sigma_{j}$ is the uncertainty at pixel $j$, and the weight $w_{j} (\lambda_{0})$ is drawn from a Gaussian:

\begin{equation} \label{eq:equation6}
\begin{split}
w_{j}(\lambda_{0}) = e^{-\frac{(\lambda_{0} - \lambda_{j})^{2}}{L^{2}}} \hspace{0.5mm} ,
\end{split}
\end{equation}

\noindent where $L$ is chosen to be 10 {{\AA}}, which is slightly larger than typical absorption features in the APOGEE spectra. This is the same procedure used in \citet{behmard2025} to continuum normalize APOGEE spectra in preparation for \emph{The Cannon}. 

The APOGEE spectra typically have high SNR values of $\sim$100/pix, and many are reported with SNR $\gtrsim$ 200/pix. These high SNR spectra have underestimated flux uncertainties. To address this, we apply the same procedure as \citet{behmard2025} and set all $<$0.005 flux uncertainty values to 0.005, as typical normalized flux uncertainties should scale as $\sim$1/SNR, and APOGEE targets have maximum effective SNR levels of $\sim$200/pix.

\subsection{Training Set}
To construct our training set for \emph{Lux}, we select late-K and M dwarfs with high-quality (SNR $>$ 50/pix) APOGEE spectra and reliable labels for stellar parameters and chemical abundances of interest. Because ASPCAP does not provide reliable abundance information for $<$4500 K stars, we cannot use them as training labels. Instead, we tag M dwarfs with the ASPCAP abundances of solar-like (F, G and early-to-mid K) binary companions, because we can assume that binary stars formed from the same parent molecular cloud and were born chemically homogeneous (e.g., \citealt{de_silva2007,de_silva2009,bland_hawthorn2010}). These solar-like ($>$4500 K) stars have reliable abundance information from ASPCAP because stellar atmosphere models work well within the solar-like temperature regime. In \citet{behmard2025}, we used a similar binary sample from APOGEE as our M dwarf training set for \emph{The Cannon}. To select binaries composed of late-K and M dwarfs with solar-like companions, we cross-match the current APOGEE data release with the \citet{elbadry2021} binary catalog. We ensure that the primaries are late-K and M dwarfs by applying the relevant type-color sequence relations from \citet{pecaut2013ApJS}, as well as a $<$4500 K on the ASPCAP-reported temperatures. To ensure that the binary companions are solar-like and have good quality abundances, we apply the following cuts on their ASPCAP parameters and abundances:

\begin{equation}\label{eq:equation9}
\begin{split}
\logg > 4 \hspace{1mm} \textrm{dex} \\
T_{\textrm{eff}} = 4500-6500 \hspace{1mm} \textrm{K} \\
[\textrm{X/H}]_{\textrm{err}} < 0.1 \hspace{1mm} \textrm{dex}
\end{split}
\end{equation}

\noindent This results in a training set of 109 binary pairs with late-K and M dwarf spectra, and labels from solar-like companions. The spectra span SNR levels of 50$-$660/pix, and the solar-like companion metallicities range from $-0.60$ $<$ [Fe/H] $<$ 0.28 dex.

\begin{figure*}[t]
    \centering
    \begin{minipage}{0.99\textwidth}
        \centering
    \includegraphics[width=0.99\textwidth]{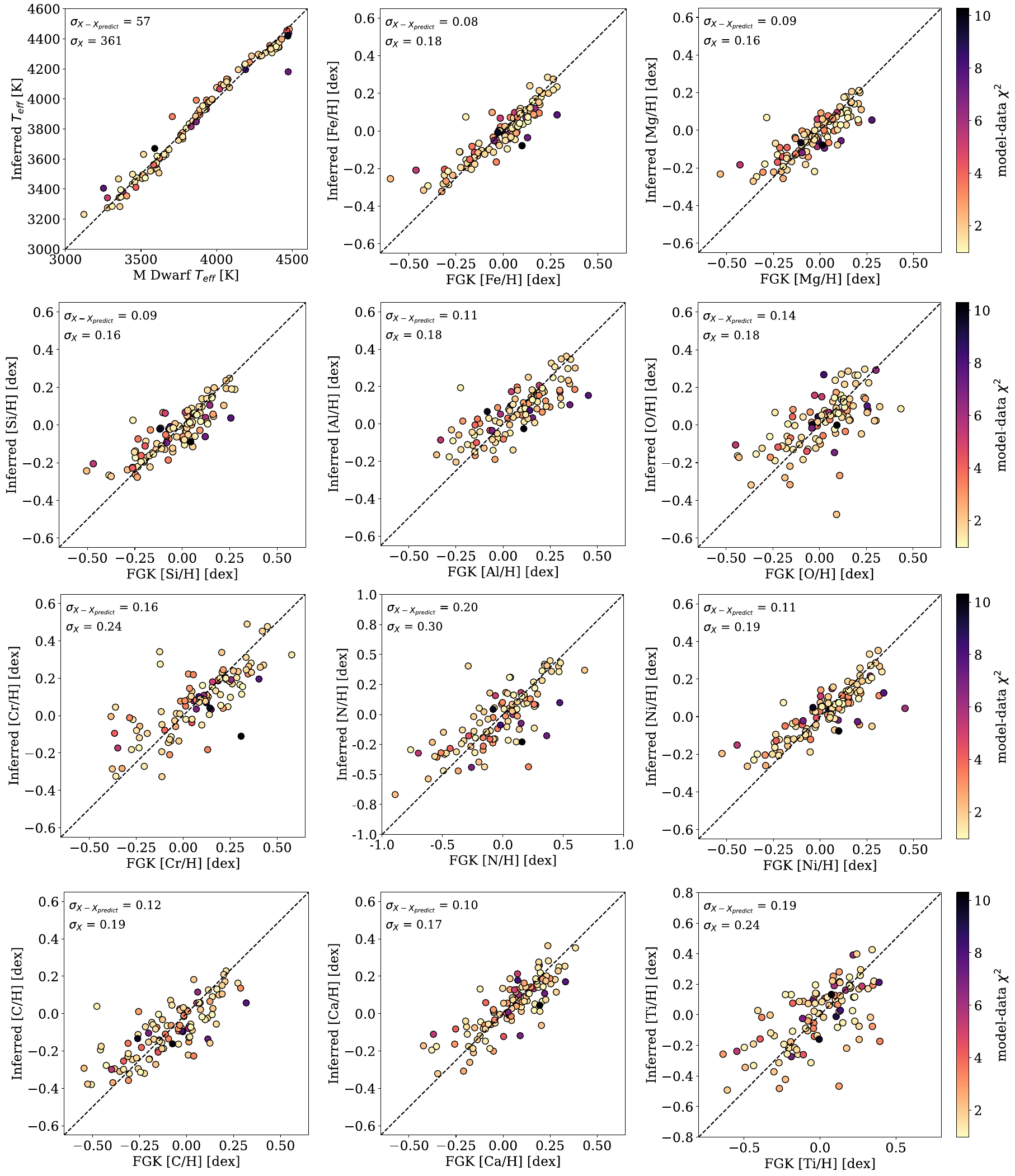} % first figure itself    
    \end{minipage} \hfill
    \caption{1-to-1 plots of our inferred vs. ASPCAP labels for our training set of late-K and M dwarfs with solar-like companions, after running a LOOCV scheme with \emph{Lux}. The points are colored by the reduced $\chi^{2}$ of the flux model spectral fit. We report the rms scatter between the inferred late-K/M dwarf and solar-like companion labels in the top left of each plot, and the scatter of the solar-like companion labels below. The former values are smaller than the latter for every label, indicating that \emph{Lux} recovers the labels using the late-K/M dwarf spectra with rms scatter that are well within the intrinsic solar-like label scatter.}
    \label{fig:figure1}
    \end{figure*}
\newpage

\begin{figure*}[t]
    \centering
    \begin{minipage}{0.99\textwidth}
        \centering
    \includegraphics[width=0.99\textwidth]{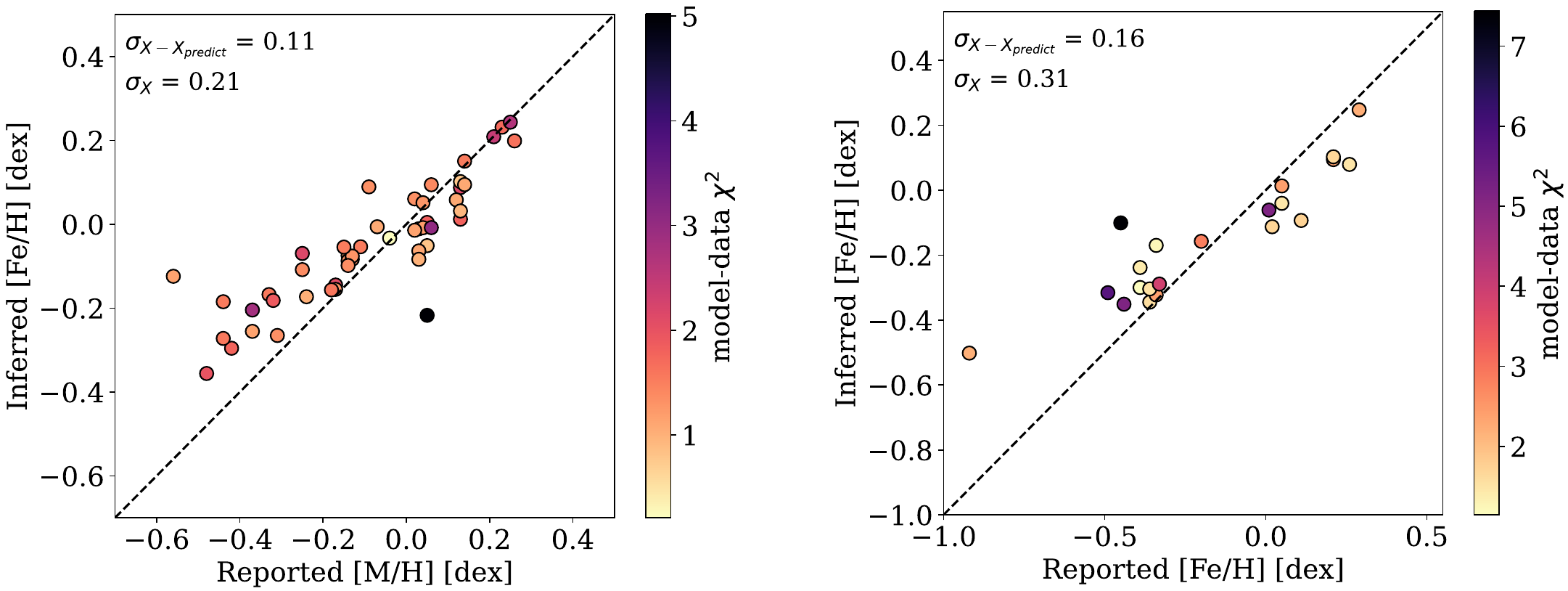} % first figure itself    
    \end{minipage} \hfill
    \caption{1-to-1 plots of our inferred [Fe/H] abundances versus the reported [M/H] and [Fe/H] abundances from the \citet{wanderley2025} (left) and \citet{souto2022} (right) samples using our \emph{Lux} model. The points are colored by the reduced $\chi^{2}$ of the flux model fit. We report the rms scatter between the inferred and reported abundances in the top left of each plot, and the intrinsic scatter of the reported abundances below. The former values are smaller than the latter for every label, indicating that our model recovers the reported abundances with rms scatter well within their intrinsic scatter.}
    \label{fig:figure2}
    \end{figure*}

Following the procedure of \citet{behmard2025}, we use $T_{\textrm{eff}}$ (late-K and M dwarf ASPCAP values) and a set of chemical abundances (Fe, Mg, Al, Si, C, N, O, Ca, Ti, V, Cr, Ni; from solar-like companions) as our training labels. As mentioned earlier, ASPCAP $T_{\textrm{eff}}$ values are considered reliable in the late-K to M dwarf regime \citep{souto2022}. In \citet{behmard2025}, we determined that this particular abundance set does the best job (among all other tested combinations of ASPCAP abundances) at reproducing the solar-like companion abundance labels that we assume to be ground truth. We test this label set for our \emph{Lux} model by running a leave-one-out cross-validation (LOOCV) scheme on our training set with the aim of reproducing the training labels. We illustrate our LOOCV results in Figure \ref{fig:figure1}, and find that we reproduce the abundance training labels to precisions of $\sim$0.08$-$0.20 dex, with the best performance for [Fe/H]. These precision values are calculated as the rms scatter between the inferred labels for late-K and M dwarfs, and the training labels from the solar-like companions. They are smaller than the rms scatter within just the solar-like companion labels (second row of values in top left corners of each Figure \ref{fig:figure1} panel), indicating that our inferred labels are learning information from the training set, and not just reproducing the training label scatter.

We remove stars in our training set with flux model spectral fit $\chi^{2}$ values $>$ 100,000 \citep{birky2020,behmard2025}. These stars have properties that are not well-represented by the rest of the training set and require additional labels to describe their spectra (e.g., stellar activity, elements not included in ASPCAP), which is beyond the scope of this study. This removes four stars, leaving 105 in our final training set. After this cut, the peak $\chi^{2}$ is $\sim$10,700, which translates to a reduced $\chi^{2}$ of $\sim$1.4 considering that there are $\sim$7400 pixels in the APOGEE wavelength range (which can be considered the number of degrees of freedom). An excellent flux model fit would yield a reduced $\chi^{2}$ of approximately one. Thus, a reduced $\chi^{2}$ of $\sim$1.4 indicates a good fit (e.g., \citealt{birky2020,rampalli2024}).

\subsection{Validation with \citet{souto2022} and \citet{wanderley2025} Datasets}

We test how well our data-driven \emph{Lux} model reproduces M dwarf abundances from stellar atmosphere modeling. Because standard stellar atmosphere models do not operate well outside of the solar-like ($\sim$4500-6500 K) regime, out-of-the-box models must be considerably modified to fit M dwarf spectra. The largest M dwarf samples with abundances from modeling APOGEE spectra with this approach are presented in \citet{souto2022} and \citet{wanderley2025}. The \citet{souto2022} sample consists of 21 M dwarfs observed by SDSS-IV/APOGEE 1 \citep{blanton2017}, with spectra collected by the APOGEE-N instrument and reduced as part of APOGEE DR16 \citep{nidever2015,ahumada2020,jonsson2020}. Starting with the APOGEE DR17 line list \citep{smith2021}, \citet{souto2022} added several prominent M dwarf spectral features (e.g., Ni I lines present in APOGEE spectra of metal-rich M dwarfs) and generated synthetic spectra with the TurboSpectrum code \citep{plez2012} and LTE MARCS model atmospheres \citep{gustafsson2008}. The synthetic spectra were manually adjusted to ensure the best fit to each APOGEE spectrum. This enabled \citet{souto2022} to robustly derive $T_{\textrm{eff}}$, \logg, Fe, C, O, Mg, Al, K, and Ca for their M dwarf sample. \citet{wanderley2025} used the same procedure to derive $T_{\textrm{eff}}$, \logg, oxygen abundances A(O), and bulk metallicity [M/H] for 48 M dwarfs with newer spectra from APOGEE DR17. 

The quality of these M dwarf abundance datasets makes them excellent validation sets for our \emph{Lux} model. We use the same implementation of \emph{Lux} as in our LOOCV test to assess how well we can reproduce reported M dwarf [Fe/H] and [M/H], respectively from \citet{souto2022} \citet{wanderley2025}. Bulk metallicity [M/H] abundances can be likened to iron abundances. We show our results in Figure \ref{fig:figure2}. Our inferred [Fe/H] values agree with the [Fe/H] and [M/H] values from \citet{souto2022} and \citet{wanderley2025} to precisions of 0.16 dex and 0.11 dex, respectively, and adhere convincingly to 1-to-1 trends. The reduced $\chi^{2}$ values of the model fit to spectra are 1.5 and 1.4 for the two samples, respectively, indicating good fits. The lower precisions in inferred [Fe/H] compared to our LOOCV test (0.08 dex) likely stem from differences in data reduction pipelines between APOGEE data releases; we use DR19 spectra for our training set, while \citet{wanderley2025} and \citet{souto2022} use DR17 and DR16 for their samples, respectively. These data reduction differences have accumulated from DR16 to DR19, so our inferred [Fe/H] precision is lower for the \citet{souto2022} sample compared to the \citet{wanderley2025} sample. Still, our [Fe/H] abundances agree well with the reported values from both these datasets, indicating that our model is robust at inferring M dwarf metallicities.

\subsection{Label Uncertainties}
To derive uncertainties on labels inferred from our \emph{Lux} model, we estimate the scatter between labels from repeat APOGEE observations of the same star. This is the same procedure used in \citet{behmard2025} to estimate label uncertainties on similar data-driven models of M dwarf spectra. We identify $\sim$500 late-K and M dwarfs that each have two very high-quality (SNR $>$ 200/pix) visit spectra that we denote $A$ and $B$. We then define a quantity $Z_{A,B}$ for each star and label:

\begin{equation} \label{eq:equation11}
Z_{A,B} = \frac{\ell_{A} - \ell_{B}}{\sqrt{2\sigma_{inflate}^{2}}}, 
\end{equation}

\noindent where $\ell_{A}$ and $\ell_{B}$ are the inferred values for a particular label included in our model ($T_{\textrm{eff}}$ or a chemical abundance) using spectrum $A$ or $B$. We fit for $\sigma_{inflate}$, which can be thought of as the additional scatter needed to account for the difference in inferred labels from different APOGEE visit spectra $A$ and $B$, and as such should include all sources of uncertainty endemic to the APOGEE data reduction pipeline. The $\sigma_{inflate}$ term is multiplied by two to account for the two visit spectra. We fit for the $\sigma_{inflate}$ factors so that the 13 distributions of $Z_{A,B}$ (corresponding to our 13 labels) from our sample of $\sim$500 stars are normal distributions centered at 0, with standard deviations of 1. The resultant $\sigma_{inflate}$ values that we take as uncertainties on $T_{\textrm{eff}}$ and chemical abundances from our model are 27.5 K and 0.03$-$0.11 dex, respectively. Our uncertainty on [Fe/H], which we use frequently throughout this study focused on M dwarf metallicity, is 0.05 dex.

\begin{figure*}[t]
    \centering
    \begin{minipage}{0.9999\textwidth}
        \centering
\includegraphics[width=0.99\textwidth]{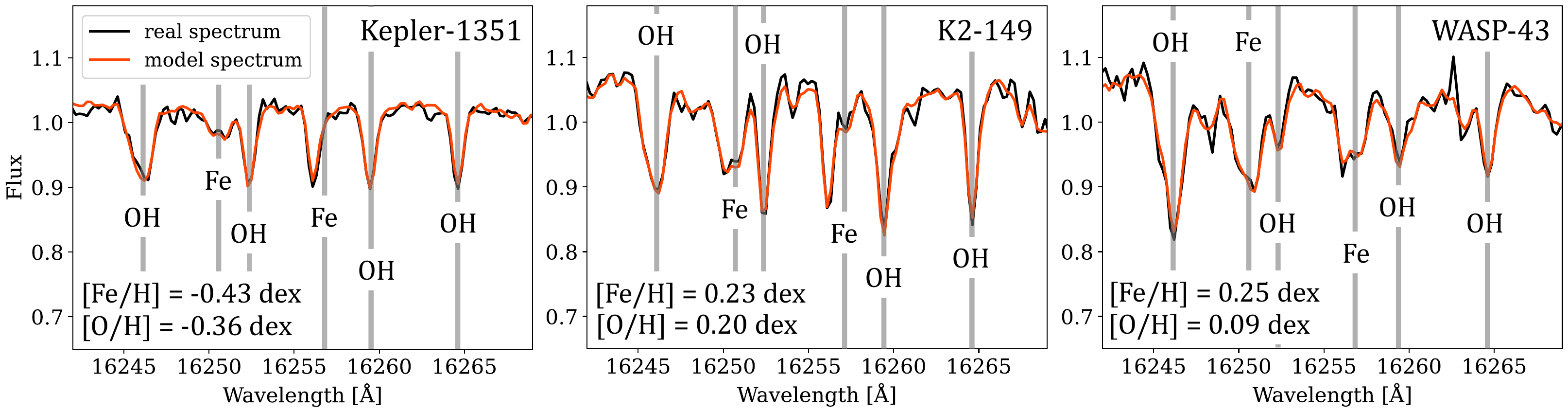} % first figure itself    
    \end{minipage} \hfill
    \caption{Plots of our flux model fits to three late-K and M dwarf planet hosts with different inferred [Fe/H] and [O/H] values (provided in the lower left corners of the panels). We highlight Fe and OH spectral features to demonstrate how our inferred abundance values are in accordance with the strengths of these lines.}
    \label{fig:figure3}
    \end{figure*}

\section{Singles vs. Multis Metallicity} \label{sec:metallicity_comparison}
\begin{figure*}[t]
    \centering
    \begin{minipage}{0.85\textwidth}
        \centering
    \includegraphics[width=0.99\textwidth]{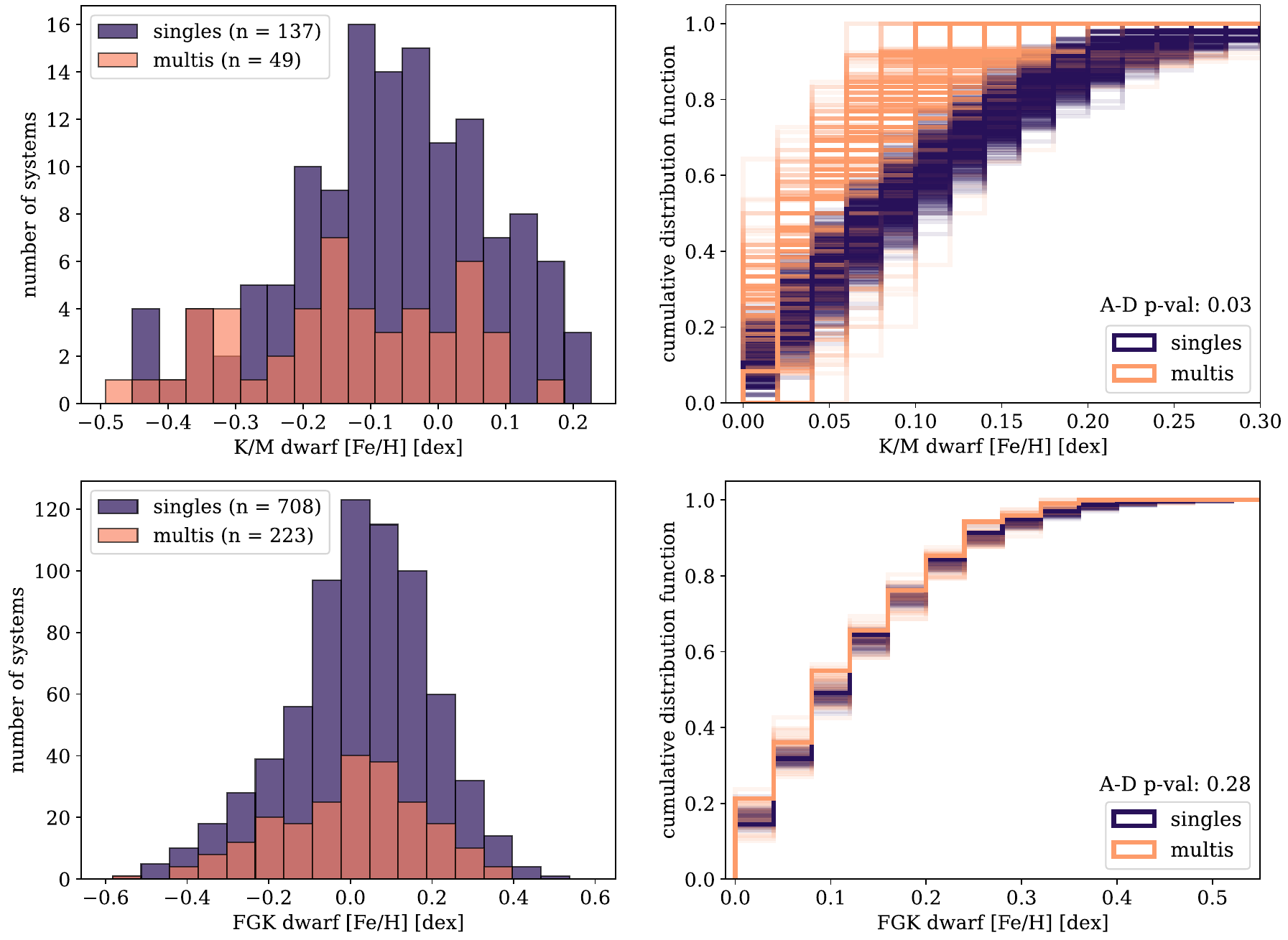} % first figure itself    
    \end{minipage} \hfill
    \caption{Histograms (left) and associated bootstrapped CDFs (right) of host star metallicities for singles (purple) and multis (orange). The top row corresponds to our total sample of late-K and M dwarf systems with metallicities, while the bottom row corresponds to our comparison sample of FGK dwarfs. The CDFs for the late-K/M dwarf sample were generated by resampling from metallicity uncertainties 1000 times, while those for the FGK dwarfs were resampled 100 times. We provide associated A-D $p$-values for singles versus multis in the lower right corners of the CDF plots.}
    \label{fig:figure4}
    \end{figure*}

\begin{figure*}[t]
    \centering
    \begin{minipage}{0.85\textwidth}
        \centering
    \includegraphics[width=0.99\textwidth]{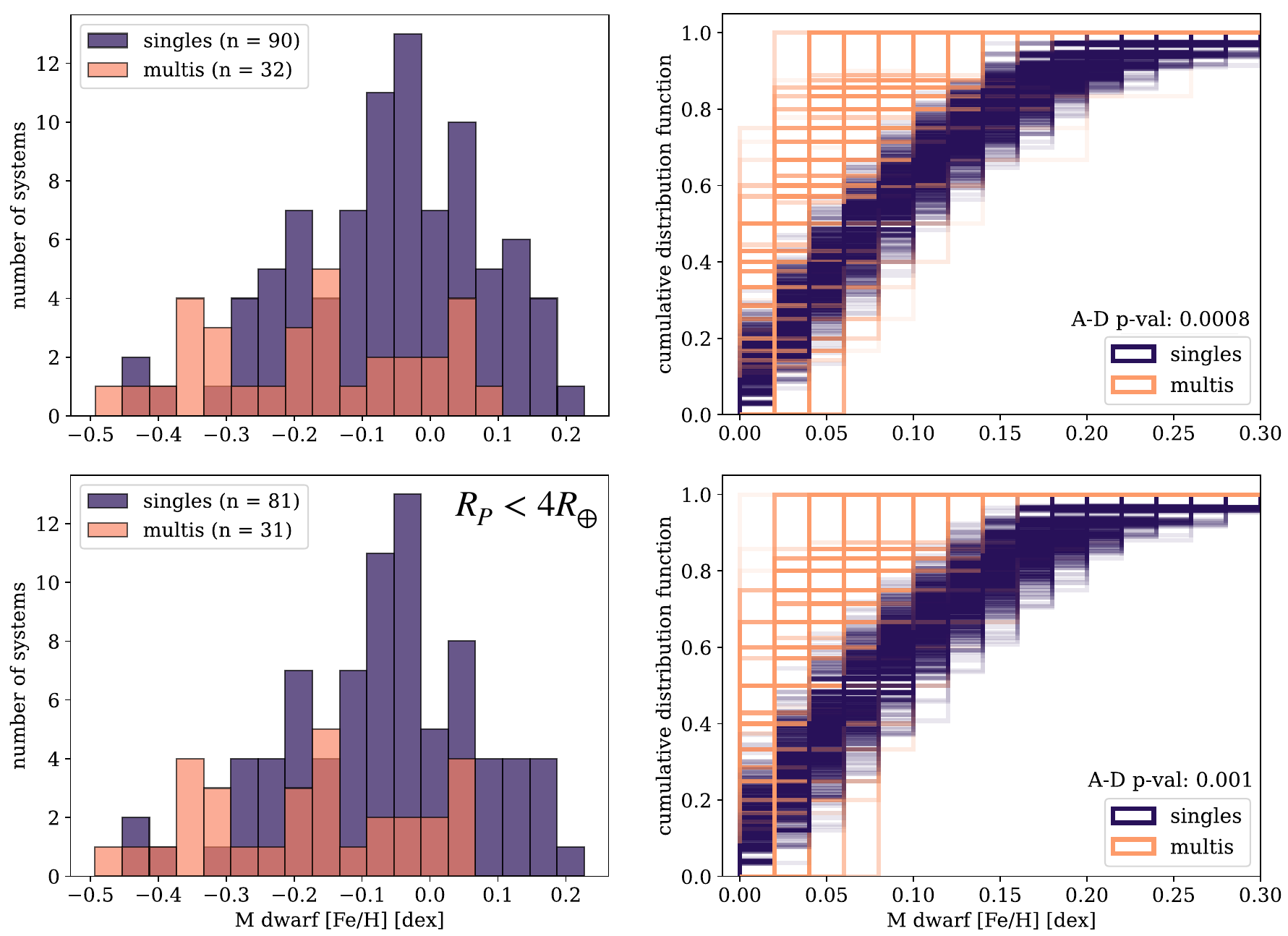} % first figure itself    
    \end{minipage} \hfill
    \caption{Histograms (left) and associated bootstrapped CDFs (right) of M dwarf host star metallicities for singles (purple) and multis (orange). The top row corresponds to our total sample of M dwarf systems with metallicities, while the bottom row corresponds to the same sample, but after removing systems with planets larger than 4 $R_{\oplus}$. We provide associated A-D $p$-values for singles versus multis in the lower right corners of the CDF plots.}
    \label{fig:figure5}
    \end{figure*}

We apply our \emph{Lux} model to our sample of late-K and M dwarfs with singles and multis to derive their chemical abundances. Specifically, this is our confirmed late-K and M dwarf planet sample with APOGEE spectra, consisting of 137 singles and 48 multis containing 125 planets. We infer our full set of \emph{Lux} model labels ($T_{\textrm{eff}}$, Fe, Mg, Al, Si, C, N, O, Ca, Ti, V, Cr, Ni), and show three example model fits in Figure \ref{fig:figure3}. We assess potential abundance differences between singles and multis. First, we examine [Fe/H], which is a proxy for bulk metallicity. We frequently refer to it as such throughout the rest of this manuscript. The metallicity distribution of late-K and M dwarfs with multis appears more metal-poor compared to that of singles (Figure \ref{fig:figure4}, upper row). To quantify this, we resample each [Fe/H] value from a normal distribution with sigma set to the metallicity uncertainty of 0.46 dex, and generate a set of 1000 metallicity cumulative distribution functions (CDFs) (similar to bootstrap analyses in, e.g., \citealt{rodriguez_martinez2023,vissapragada2025}). The resulting distributions appear quite different between singles and multis. We then perform a two-sample Anderson-Darling (A-D) test on each of the 1000 pairs of resampled CDFs, and calculate a mean $p$-value of 0.032, i.e., moderate evidence that the metallicity distributions of singles and multis do not belong to the same underlying population. If we restrict to M dwarfs ($<$4000 K), our planet sample gets smaller (90 singles and 32 multis containing 83 planets), but the difference in metallicity CDFs of singles and multis appears more significant (Figure \ref{fig:figure5}, upper row). The associated A-D $p$-value is 0.0008, which constitutes strong evidence that M dwarfs with multis are significantly more metal-poor compared to those with singles. Tentative evidence ($p$ = 0.015) for this result was first reported in \citet{anderson2021} using indirect metallicity proxies. More recently, \citet{rodriguez_martinez2023} and \citet{wanderley2025} reported it with significance ($p$ = 0.001 and 0.002, respectively) using samples of heterogeneous metallicities (i.e., derived from different abundance pipelines), and homogeneous metallicities for $<$50 systems, respectively. We use our larger sample of homogeneous abundances to further confirm this result at higher significance. We find that host star metallicity decreases with planet multiplicity, which is also reported in \citet{rodriguez_martinez2023}. For our systems with 1, 2, 3, and 4$+$ planets, the average [Fe/H] value is $-$0.05, $-$0.13, $-$0.21, and $-$0.29 dex, respectively.

%but at lower significance levels, likely because their metallicity samples are heterogeneous (i.e., derived from different abundance pipelines) and smaller, respectively. 

Many studies have demonstrated that giant planets are preferentially found around metal-rich host stars. This is well-known among FGK dwarfs (e.g., \citealt{santos2004,fischer2005}) and more recently was demonstrated for M dwarfs as well \citep{gan2025}. Giant planets are also preferentially found in single systems where they are the only observed planet (e.g., \citealt{steffen2012}). To re-assess our M dwarf singles versus multis results in this context, we restrict our sample to small planets ($<$4 $R_{\oplus}$). This cuts our sample to 81 singles and 31 multis containing 81 planets. This is not much smaller than our sample including giant planets because most of the planets in our sample are small. We apply the same analysis and generate 1000 pairs of metallicity CDFs for singles and multis (Figure \ref{fig:figure5}, lower panel), and find that the resulting A-D $p$-value decreases in significance, but is still strong ($p$ = 0.001). This demonstrates that the metallicity difference holds for small planets and is independent of known trends between giant planets, their proclivity for single-planet systems, and host star metallicity.

%This is potentially consistent with the results of \citet{brewer2018b}, who found that compact multis (systems with 3$+$ planets that orbit within 1 AU) around FGK dwarfs are more metal-poor compared to singles.

\subsection{Comparison with FGK Dwarf Systems}
Removing late-K dwarfs from our sample and restricting to M dwarfs results in a stronger metallicity difference between singles and multis. This suggests that the difference depends on stellar type, which potentially agrees with previous studies that report no discernible difference in metallicities between singles and multis around FGK dwarfs (e.g., \citealt{Romero2018,weiss2018}). We assess this with a comparison sample of singles and multis around FGK dwarfs (4500 $<$ $T_{\textrm{eff}}$ $<$ 6500 K and \logg \hspace{1mm}$>$ 4 dex) with high-quality APOGEE spectra (SNR $>$ 50/pix) and reliable abundances ([X/H]$_\textrm{err}$ $<$ 0.1 dex) from the ASPCAP pipeline. We construct this sample by cross-matching transiting planets in the Exoplanet Archive with SDSS-V DR19 after applying the cuts described above to ASPCAP quantities. This yields a sample of 708 singles and 223 multis hosting 563 planets. We compare the metallicity CDFs of singles and multis using their [Fe/H] values from ASPCAP (Figure \ref{fig:figure4}, lower row). Because the ASPCAP pipeline is reliable for solar-like stars, we can use the FGK dwarf ASPCAP abundances directly (which was not possible for our M dwarf sample). We resample from the metallicity uncertainties 100 times and recover an A-D $p$-value of 0.29, indicating that the metallicity distributions of multis and singles around FGK dwarfs are not statistically distinguishable. This is in agreement with previous studies (\citet{Romero2018} and \citet{weiss2018} report $p$-values of 0.43 and 0.29, respectively), and suggests that the apparent trend of multis orbiting more metal-poor stars compared to singles emerges in the transition from the FGK to M dwarf regime.

%Mention that you see an increase in planet size with host star metallicity among both the singles and multis, which is what we expect given that the planet-metallicity correlation is shown to be true for M dwarfs.

\begin{figure*}[t]
    \centering
    \begin{minipage}{0.85\textwidth}
        \centering
    \includegraphics[width=0.99\textwidth]{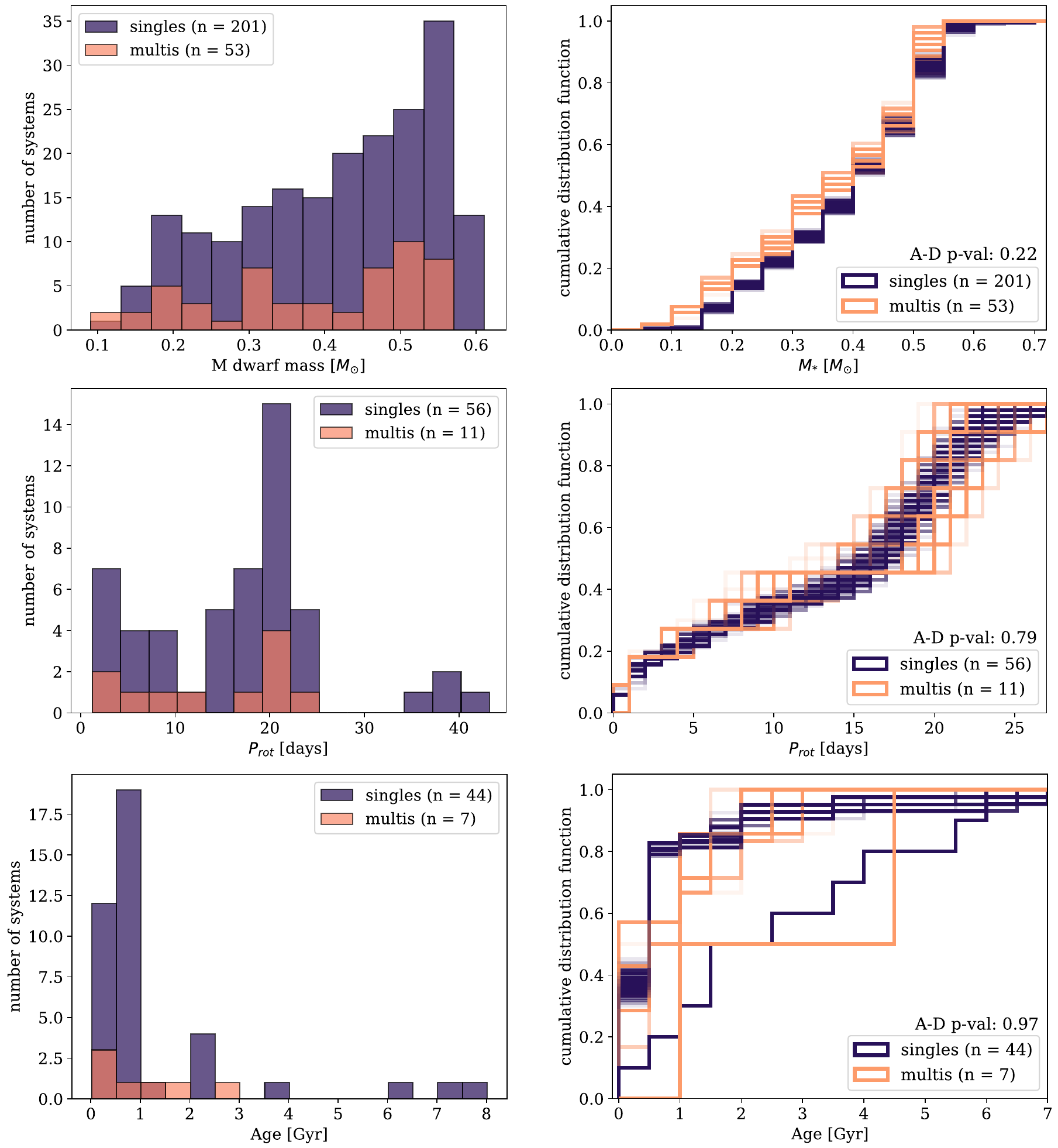} % first figure itself    
    \end{minipage} \hfill
    \caption{Histograms (left) and associated bootstrapped CDFs (right) of M dwarf host star properties (masses, rotation periods, and ages) for singles (purple) and multis (orange). The top row corresponds to empirically-determined host star masses, the middle row to rotation periods derived from photometric measurements, and the bottom row to ages measured from gyrochronology. We provide associated A-D $p$-values for singles versus multis in the lower right corners of the CDF plots.}
    \label{fig:figure6}
    \end{figure*}

\subsection{Chemical Abundances Beyond Bulk Metallicity}
Our \emph{Lux} model provides chemical abundances beyond [Fe/H]. We use these abundances of other elements (Mg, Al, Si, C, N, O, Ca, Ti, V, Cr, and Ni) to explore potential trends between M dwarf singles and multis in more detailed chemical space. We find that M dwarfs with multis are significantly more metal-poor compared to those with singles for all elements in [X/H] form, which is unsurprising given that all abundances are correlated with each other through nucleosynthesis. The A-D $p$-values from comparing singles versus multis for all [X/H] chemical abundances range from $p$ = 0.0002$-$0.02. 

We also test if there are trends between singles and multis if we remove the bulk metallicity contribution (i.e., in [X/Fe] form). Before resampling from the abundance uncertainties via our bootstrap approach, there are significant differences in several $\alpha$-abundances between multis and singles; [Mg/Fe], [Si/Fe], and [O/Fe] have A-D $p$-values that range from $p$ = 0.01$-$0.0002. Multis are enhanced compared to singles in all cases. Once we apply the bootstrap approach, only [O/Fe] remains significant ($p$ = 0.016). However, this changes again if we restrict to small planets ($<$4 $R_{\oplus}$) to remove the effect that singles are often giant planets hosted by metal-rich stars. Bootstrap analysis considering only small planets yields more significant trends ($p$ = 0.15, 0.002, and 0.0009 for [Mg/Fe], [Si/Fe], and [O/Fe], respectively). Thus, it is possible that larger amounts of $\alpha$-elements in low-Fe environments enhances formation of multiple small planets over single small planets.

The A-D $p$-values of [N/Fe] and [Cr/Fe] for M dwarf singles versus multis are also quite significant ($p$ = 0.0002 and 0.001, respectively). Unlike trends with the $\alpha$-elements in our sample, [N/Fe] and [Cr/Fe] are depleted in singles compared to multis. These elements are produced via different nucleosynthetic channels, with Cr predominantly from Type la supernovae as an Fe-peak element, and N through the CNO cycle. This makes it difficult to identify a single astrophysical mechanism behind these trends. We conclude that a larger sample of M dwarf planet hosts with detailed chemical abundance information is needed to identify differences between singles and multis beyond bulk metallicity.

\section{Additional Host Star Properties} \label{sec:additional_star_properties}
To test the source of the metallicity discrepancy between M dwarf hosts of singles and multis, we examine other host star properties such as mass, rotation period, and age. 

\subsection{Mass}
Previous studies do not report significant differences in host star mass between FGK dwarfs that host singles versus multis (e.g., \citealt{weiss2018}). We use our sample to test if this holds in the M dwarf regime. To obtain homogeneous masses for our M dwarf hosts, we use the empirical relation between M dwarf mass and Two Micron All Sky Survey (2MASS) $M_{K_{S}}$-band magnitude reported in \citet{mann2019}. We first assess our larger sample that is not cross-matched with SDSS-V DR19 that contains 370 singles and 248 planets within 97 multis around late-K and M dwarfs. We are solely interested in the M dwarf regime, so we remove late-K dwarf systems from our sample by imposing the type-color relation for M dwarfs from \citet{pecaut2013ApJS}. This leaves us with 201 singles and 136 planets within 53 multis. As in our chemical abundance analysis, we resample each host star mass from a normal distribution with sigma set to its associated uncertainty value 1000 times. We then compare the host star mass CDFs between our singles and multis, and find them to be statistically indistinguishable with an A-D $p$-value of 0.22 (Figure \ref{fig:figure6}, top row). We conclude that there is no significant difference in host star mass between singles and multis around M dwarfs, similar to findings for FGK systems.
    
\subsection{Rotation Period}
While FGK dwarfs that host singles versus multis do not appear to have significantly different rotation periods (e.g., \citealt{weiss2018}), this may differ for M dwarf hosts. Several studies find this: \citet{ballard2016} report that M dwarf multis orbit slightly faster rotators than singles, while \citet{rodriguez_martinez2023} report the opposite. In both cases, the differences in rotation period between singles and multis are only marginally significant ($\sim$2$\sigma$). We investigate this further with our M dwarf sample by measuring rotation periods with observations from photometric surveys such as \emph{Kepler} \citep{Borucki2010}, TESS \citep{ricker2015}, and ZTF \citep{Masci2019}. We cross-match our sample with existing rotation period catalogs \citep{McQuillan2014, Santos2019, Irwin2011, Newton2017, Berta2012, Lu2022, Lu2024a, Lu2024b, Boyle2026}, which results in 56 singles and 24 planets within 11 multis with reported rotation periods. We assume uncertainties of 10\%, the nominal accepted value for differential rotation measurements \citep{epstein2014}. Again, we resample from the rotation period uncertainties 1000 times to generate CDFs for our multis versus singles. We find them to be statistically indistinguishable with an A-D $p$-value of 0.79, and conclude that our M dwarf hosts of multis versus singles do not have significantly different rotation periods (Figure \ref{fig:figure6}, middle row).

\subsection{Age}
We use stellar rotation periods and photometric effective temperatures derived from extinction-corrected \emph{Gaia} DR3 BP-RP colors to estimate ages via gyrochronology \citep{Barnes2003}. Gyrochronology is currently the most widely applicable age-dating method for single main-sequence stars; aside from stellar rotation, most observable stellar properties evolve only slightly during the main-sequence phase. We use the extinction map taken from the \texttt{Bayestar19} 3D dust map \citep{Green2019}, and the gyrochronology relation from \texttt{GPgyro} \citep{Lu2024a}\footnote{Available at https://github.com/lyx12311/GPgyro}, which is an empirical gyrochronology relation calibrated using Gaussian Processes on kinematic ages. To derive gyrochronology ages for our M dwarf sample, we must restrict to systems with reliable rotation periods and photometric temperatures. This reduces our sample to 44 singles and 15 planets within 7 multis. Our bootstrap approach of resampling 1000 times from age uncertainties yields an A-D $p$-value of 0.97, indicating no significant difference (Figure \ref{fig:figure6}, bottom row). However, this is a small sample with few multis. We conclude that more systems are needed to robustly assess age. 

%I think it could also be interesting to mention how there is no difference in orbital period of the planets across the fully convective boundary (Jao gap) since fully convective stars might have a different dynamo compared to partially convective stars, which might affect their orbital periods differently.

\begin{figure*}[t]
    \centering
    \begin{minipage}{0.85\textwidth}
        \centering
    \includegraphics[width=0.99\textwidth]{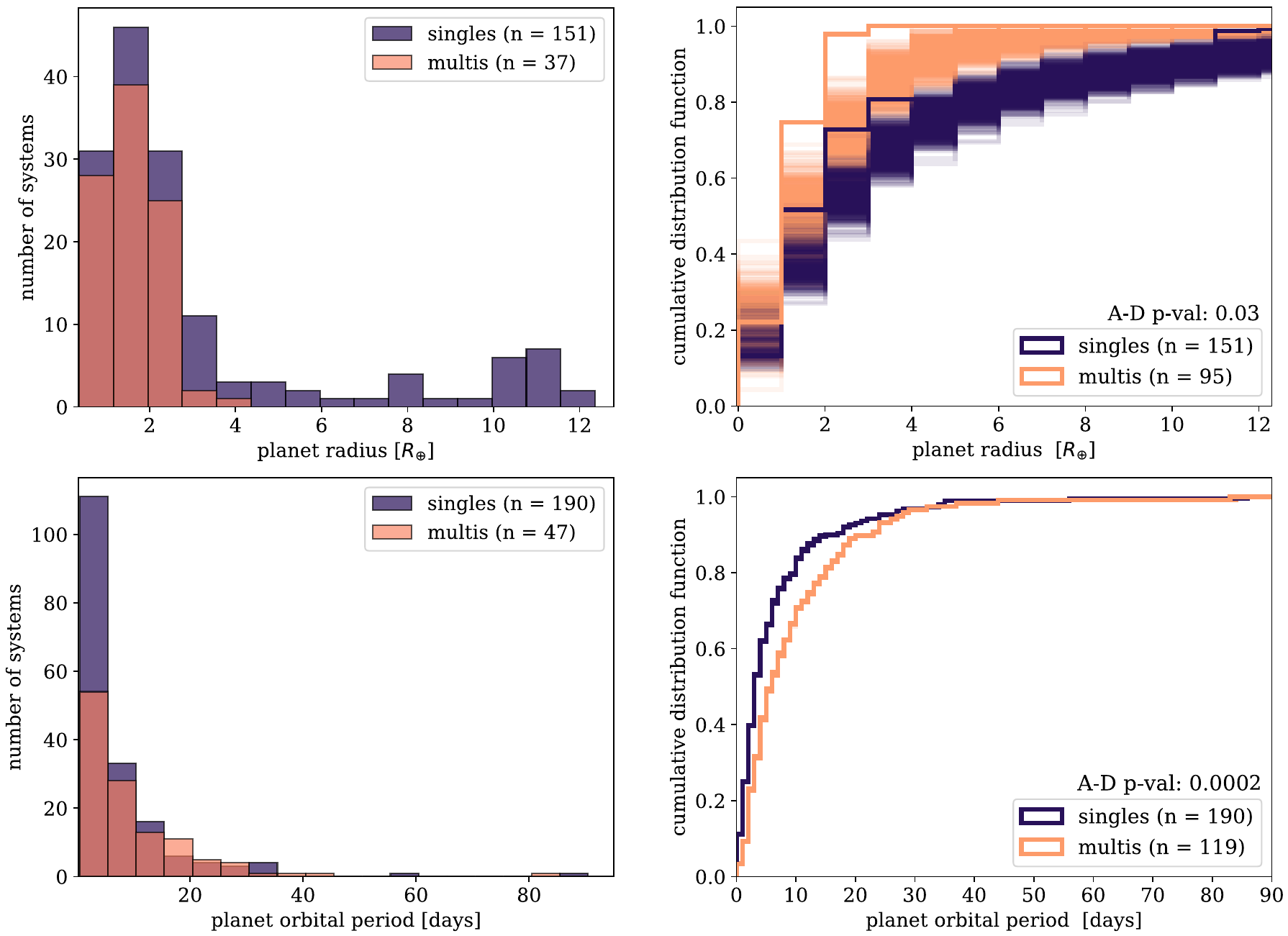} % first figure itself    
    \end{minipage} \hfill
    \caption{Histograms (left) and associated bootstrapped CDFs (right) of planet radii (top) and orbital periods (bottom) for singles (purple) and multis (orange). We provide associated A-D $p$-values for M dwarf singles versus multis in the lower right corners of the CDF plots.}
    \label{fig:figure7}
\end{figure*}

\section{Planet Properties} \label{sec:planet_properties}
We do not uncover any significant differences in host star properties beyond metallicity between M dwarf singles and multis. This suggests that the host star metallicity difference stems from planet formation rather than aspects of the host star. We examine this by comparing planet properties of singles versus multis here. 

\subsection{Planet Radius}
We first examine planet radii between M dwarf singles and multis. Because this analysis does not involve chemical abundances, we begin with our full M dwarf sample of 201 singles and 53 multis that contain 136 planets (as opposed to our smaller sample cross-matched with SDSS-V DR19). We also add a cut of $<$30\% fractional uncertainty on $R_{p}/R_{*}$ values from the Exoplanet Archive. This leaves us with 151 singles and 37 multis containing 95 planets. To compare planet radii robustly, we require homogeneous measurements. We re-derive planet radius values from the reported $R_{p}/R_{*}$ values for each planet, and homogeneous host star radii. We calculate host star radii using the empirical relation between M dwarf radii and $M_{K_{s}}$ absolute magnitude established in \citet{wanderley2025b} with M dwarfs observed in APOGEE DR17:

\begin{equation}
R_{*} = a_{0} + a_{1}M_{K_{s}} + a_{2}M_{K_{s}}^{2}
\end{equation}

\noindent where $a_{0}$ = 1.7420, $a_{1}$ = $-$0.2925, and $a_{2}$ = 0.0123, and their standard deviations are, respectively, 0.1599, 0.0522, and 0.0042. We incorporate these uncertainties as well as uncertainties on $M_{K_{s}}$ by sampling from Gaussian distributions with 1$\sigma$ set as the uncertainty values. 

We apply our bootstrap approach of resampling from planet radius uncertainties 1000 times. We find that singles preferentially host giant planets ($R_{p}$ $>$ 6 $R_{\oplus}$), which is a known trend among FGK and M dwarf systems (e.g., \citealt{steffen2012,rodriguez_martinez2023}). This trend creates a slight difference in planet radii between singles and multis (A-D $p$ = 0.030) (Figure \ref{fig:figure7}, top panel). This is similar to planet radii results for FGK dwarf singles and multis; \citet{weiss2018} report an A-D $p$-value of 0.02. If we restrict our sample to small planets ($R_{p}$ $<$ 4 $R_{\oplus}$), our A-D $p$-value increases to 0.45, indicating no significant difference. This is again similar to results for FGK dwarf systems from \citet{weiss2018}, who report an A-D $p$-value of 0.44 after the same cut. We conclude that singles and multis around M dwarfs exhibit no significant differences in planet radii, aside from the known trend that giant planets are preferentially found in single systems. These results are indistinguishable from those of singles and multis around FGK dwarfs.

\subsection{Planet Orbital Period}
We next examine planet orbital periods between singles and multis. As in our planet radius analysis, we begin with our full M dwarf sample of 201 singles and 53 multis containing 136 planets. We remove planets that lack orbital period measurements from the Exoplanet Archive, leaving 190 singles and 47 multis containing 119 planets. There are significantly more singles at short orbital periods ($P$ $<$ 5 days) compared to multis (Figure \ref{fig:figure7}, bottom panel). Beyond 19 days, the orbital periods of singles and multis are similar. These trends are also observed in singles and multis around FGK dwarfs (e.g., \citealt{weiss2018}) and other M dwarf samples \citep{rodriguez_martinez2023,wanderley2025}.

We apply our bootstrap approach of resampling from orbital period uncertainties 1000 times. This yields an A-D $p$-value of 0.0002 (Figure \ref{fig:figure7}, bottom panel), indicating that the planets in M dwarf singles have significantly shorter orbital periods compared those in multis. If we restrict to small ($R_{p}$ $<$ 4 $R_{\oplus}$) planets, the A-D $p$-value remains significant at 0.0004. This is similar to reported differences in orbital period between FGK singles and multis ($p$ = 0.002, \citealt{weiss2018}), though our result is an order-of-magnitude stronger. However, our sample is smaller and may be subject to more observational biases. The mechanism behind why single planets tend to cluster at $<$5 day orbital periods for FGK and M dwarf systems is likely the same. 

The excess of short-period single planets is likely shaped by dynamical evolution \citep{steffen2016}. For example, these planets may have originated in multi-planet systems that migrated inward in resonant chains. The outcome is thought to be a single short-period inner planet, and additional planets at periods beyond 10 days (e.g., \citealt{rodriguez_martinez2023}). The longer period planets might be dynamically excited to large mutual inclinations, making them undetectable via transits. A similar outcome can be reached in the absence of resonances, for example through planet-planet scattering (e.g., \citealt{chatterjee2008}, or chaotic secular interactions between planetary companions \citep{petrovich2019}. These dynamical scenarios are consistent with observations that systems with ultra-short period planets (USPs, $P$ $<$ 1 day) and additional transiting planets have wider period ratios between their planets when compared to multi-planet systems without USPs (e.g., \citealt{sanchis_ojeda2014}). They are also consistent with observations demonstrating that multi-planet systems with the shortest-period planets ($a/R_{*}$ $<$ 5, or $P$ = 1.3 days for a Sun-like star) are also those with the largest mutual inclinations between other planets in the system \citep{dai2018}. These results indicate that our observed excess of short-period planets in M dwarf singles is the outcome of similar dynamical effects.

\subsection{Planet Eccentricity}
It is well-established that multi-planet systems are less eccentric compared to single planet systems around FGK dwarfs (e.g., \citealt{limbach2015}). This was recently demonstrated for M dwarf systems as well \citep{sagear2023}. Thus, singles and multis are characterized by distinctly hotter and cooler dynamics, respectively. We test if this eccentricity effect can be seen in our sample of M dwarf systems with metallicities derived from APOGEE observations. To measure orbital eccentricities for our sample cross-matched with SDSS-V DR19, we leverage the photoeccentric effect \citep{Dawson_2012}. Photoeccentricities rely on precise characterization of the ingress and egress from transit light curves, and we use precise transit light curves from \textit{Kepler} and TESS for our analysis. Restricting our sample to systems with \textit{Kepler} or TESS data yields 28 TESS singles, 7 TESS multis systems containing 13 planets, 14 Kepler singles, and 9 Kepler multis systems containing 20 planets. This totals 28 singles and 33 multis with measurable photoeccentricities. We exclude systems with only K2 observations because the nature of K2 data is such that ingress and egress shapes can be characterized much less precisely than for \emph{Kepler} and TESS. We expect that excluding K2 targets at this sample size will not measurably impact the precision of the underlying eccentricity distribution.

We summarize the transit fitting process here; the full process, including transit fit parameters and prior distributions, is described in detail in \citet{Sagear_kinematic_2026}. We detrend and fit the \textit{Kepler} and TESS transit light curves with the ALDERAAN transit fitting pipeline \citep{gilbert2025}. Light curves are uniformly detrended and processed within the pipeline, with transit timing variations incorporated where necessary for \textit{Kepler} systems, and nested sampling is used to sample the high-dimensional transit posteriors. We extract $e$ and $\omega$ distributions from the transit posteriors using post-model importance sampling \citep{MacDougall_2023}. 

The precision of stellar host density measurements directly affects the precision of eccentricity measurements from photometry. For the \emph{Kepler} sample of planet hosts, we use the \citet{Mann_2015} and \citet{Mann_2019} empirical stellar density formulas, leveraging stellar parallax measurements from \emph{Gaia} \citep{gaia_2016,gaia_dr3} along with $K_s$ magnitudes from 2MASS \citep{Skrutskie_2006}. For the TESS sample, we use stellar densities published in the TESS Input Catalog's list of known Cool Dwarf Targets \citep{muirhead2018, Stassun_2018}. These sources provide precise stellar densities for M dwarfs than stellar properties from isochrone models, which tend to be poorly modeled in this mass regime. We infer the underlying eccentricity distributions for the population of singles and multis within a Bayesian hierarchical framework; full details of this process are described in \citet{Sagear_radius_2026} and \citet{Sagear_kinematic_2026}. We take the underlying eccentricity distribution to resemble a Rayleigh distribution and apply a prior on the Rayleigh sigma parameter of $\sigma_R \sim U(0,1]$. The geometric transit probability is non-uniform as a function of eccentricity \citep{kipping_bayesian_2014}; we take care to account for this as a prior within our Bayesian framework for transiting planets. 

Individual planet photoeccentricities are inherently imprecise, in part because the eccentricity measurement is highly degenerate with the system viewing angle (or longitude of periastron $\omega$). We include eccentricity measurements from RV observations, if available. Among the systems in our sample, 12 singles and 3 multis containing 7 planets have reliable RV data and posterior measurements. The planets with eccentricity measurements from RV observations and their sources are as follows: L 98-59 b, c, and d \citep{Cadieux_2025}, HD 260655 b and c \citep{Luque2022}, TOI-4342 b and c \citep{Parc_2026}, K2-18 b \citep{Sarkis_2018}, Kepler-16 b \citep{triaud_2022}, Kepler-45 b \citep{Johnson_2012}, TOI-1452 b \citep{Cadieux_2022}, TOI-1266 b \citep{Greklek-McKeon_2025}, Ross 176 b \citep{Ross176}, TOI-654.01 \citep{ikuta_2025}, L 168-9 b \citep{Hobson_2024}, TOI-269 b \citep{Cointepas_2021}, K2-25 b \citep{Stefansson_2020}, TOI-3785 b \citep{Powers_2023}, TOI-1728 b \citep{Kanodia_2020}. If full eccentricity posteriors from RV data are not available, we take the published eccentricity mean and uncertainty and assume the eccentricity distribution is a truncated normal distribution between 0 and 1.

\begin{figure}[t]
    \centering
    \includegraphics[width=\linewidth]{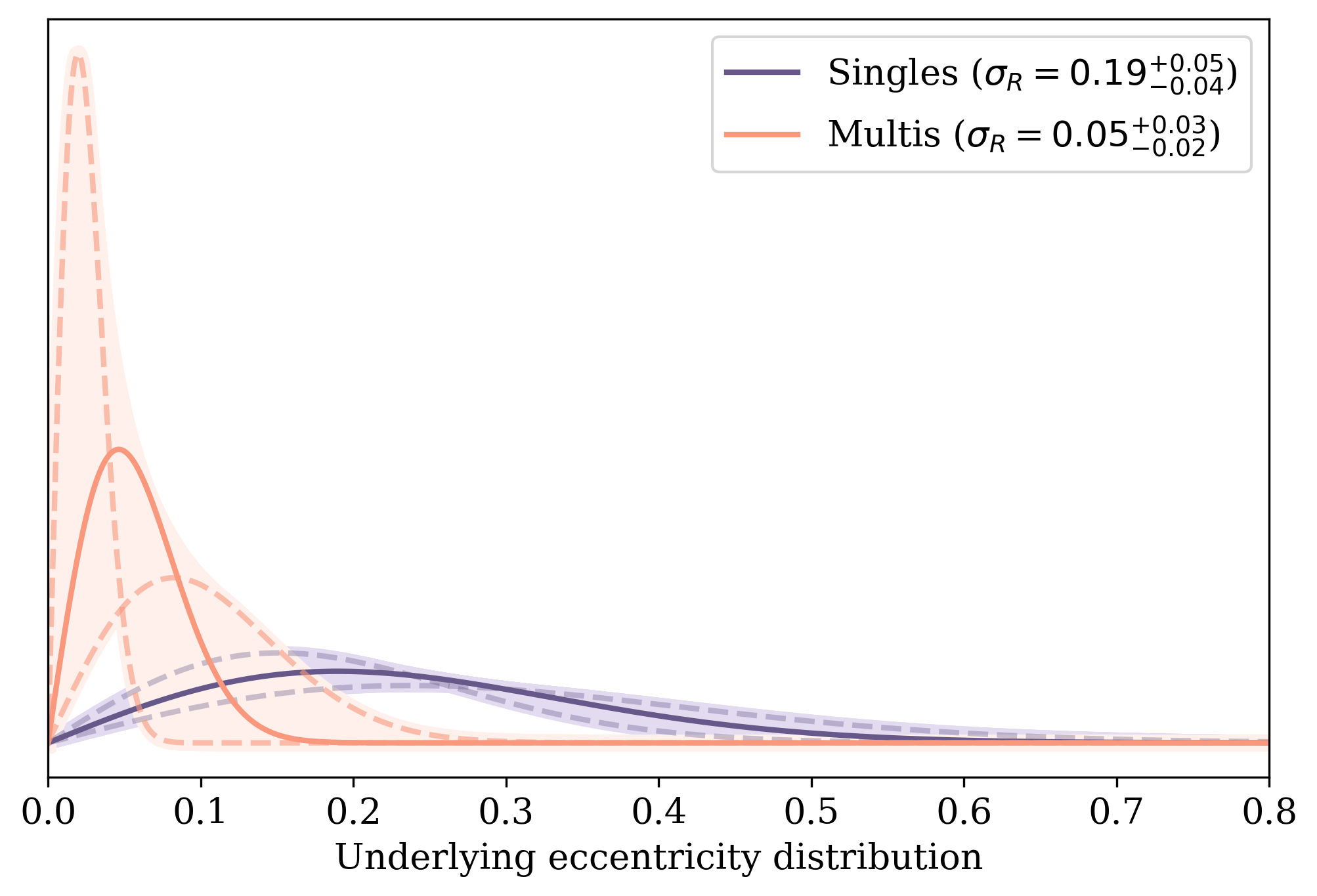}
    \caption{Underlying eccentricity distributions for M dwarf singles and multis. Photoeccentricities are derived from \emph{Kepler} and TESS data, and are combined with eccentricities from RV measurements where available.}
    \label{fig:eccentricities}
    \end{figure}

We hierarchically model the underlying eccentricity distribution of both singles and multis as Rayleigh distributions and present the best-fit models and their uncertainties, shown in Figure \ref{fig:eccentricities}. The underlying eccentricity distributions for this sample of singles and multis are consistent with the results of \citet{sagear_ballard}, who conducted a similar analysis using a homogeneous sample of photoeccentricities for M dwarf planets with \emph{Kepler} data. Like \citet{sagear_ballard}, we conclude that planets in M dwarf singles are significantly more eccentric than those in M dwarf multis.

\newpage

\section{Discussion} \label{sec:discussion}
We find strong evidence that M dwarf hosts of multiple transiting planets (``multis") are significantly more metal-poor than those with only one detected transiting planet (``singles") ($p$ = 0.0008). This trend holds if we consider only small ($<$4 $R_{\oplus}$) planets ($p$ = 0.001), and is thus independent of the planet-metallicity correlation, and the known trend that giant planets are often ``lonely" (i.e., found in singles). We do not uncover significant differences in any other host star property tested here (mass, rotation period, or age) between singles and multis. However, we do find differences in planet properties, where planets in singles exhibit shorter orbital periods and higher eccentricities compared to those in multis. This suggests that metallicity differences in M dwarf systems instigate different planetary dynamics that give rise to different system architectures, and ultimately the distinction between singles and multis that we observe.

There are many dynamical pathways linked to metallicity that may sculpt singles as opposed to multis. For example, higher metallicity systems are more likely to produce giant planets according to the planet-metallicity correlation. These giants may dynamically perturb initially co-planar planetary companions as they migrate in. For example, a giant planet that forms at the outskirts of a metal-rich system may initiate planet-planet scattering during migration, potentially ejecting planets from the system, or tilting their orbits to high mutual inclinations that prevent them from transiting the host star \citep{chatterjee2008}. Alternatively, dynamical stirring could be instigated in the absence of giant perturbers by inward migration of small planets in resonant chains \citep{izidoro2017}, or secular chaos \citep{petrovich2019}. Such dynamically active scenarios are favored in higher metallicity systems that can form more planets with heavier cores on shorter timescales \citep{dawson2013}. A likely outcome in all cases is only one transiting planet left behind. In accordance with our results, this planet may exhibit evidence of an eventful dynamical past, such as a high orbital eccentricity and/or a short orbital period resulting from inward migration. 

There is substantial observational evidence that these dynamical scenarios occur in both FGK and M dwarf systems. As mentioned earlier, giant planets are often found as the lone planet in single systems, and preferentially form around metal-rich FGK and M dwarfs (e.g., \citealt{gan2025,bryan2025}). Planets in singles also have higher orbital eccentricities compared to those in multis with both FGK and M dwarf hosts \citep{limbach2015,sagear2023}. Recently, it was demonstrated that small planets with high mutual inclinations are more common around metal-rich stars of both types \citep{hua2025}. 

Our discovery that M singles are more metal-rich than multis fits within this framework, as does our finding that host star metallicity decreases with planet multiplicity in multis (which indicates a more dynamically quiet history). 
%Notably, these results hold when we restrict to small planets ($R_{p} < 4$ $R_{\oplus}$) and thus remove the dependence on known trends concerning giant planets and metallicity. 
What is surprising is that these results are not echoed in FGK dwarf systems. \citet{Romero2018} was the first to demonstrate a lack of metallicity differences between FGK dwarf hosts of singles and multis from the \emph{Kepler} mission. Later, \citet{weiss2018} examined a larger sample of FGK systems with homogeneously-measured host star and planet properties, and found no significant differences between singles and multis for any of them, including metallicity ($p$ = 0.29). They interpreted this as evidence that singles and multis are drawn from the same underlying planet population. We find a near-identical result for the FGK singles and multis in our sample; they do not exhibit a significant difference in metallicity ($p$ = 0.29). Nor does our M dwarf sample if we include late-K dwarf systems, though evidence for a tentative difference begins to emerge ($p$ = 0.032). Only when we restrict to M dwarf systems does the difference become significant ($p$ = 0.0008), indicating that it arises at the late-K to M dwarf transition. 

Why do observe a metallicity difference between singles and multis around M dwarfs, but not FGK dwarfs? The answer may be related to protoplanetary disk mass. M dwarfs are less massive than FGK dwarfs, and accordingly host less massive disks with smaller amounts of solid, planet-forming material. In the M dwarf case, metallicity may be the deciding factor in whether the planetary mass budget is large enough to instigate dynamical evolution that leaves one transiting planet behind. In contrast, comparatively massive disks around FGK dwarfs may often contain enough solid material to form many planets with massive cores on short timescales that instigate disruptive dynamics, even in the low metallicity regime. 

Other possibilities are related to protoplanetary disk size. M dwarf disks are smaller than FGK dwarf disks, with a smaller radial extent. This implies that planets which form from an M dwarf disk will all reside relatively close to the host star, compared to those from an FGK dwarf disk. Close proximity of planets in the M dwarf case is more likely to instigate dynamics that result in a single transiting planet. This is enhanced in the high metallicity case that favors formation of giant planets, which are more likely to wreak dynamical havoc. Another explanation related to disk extent is that multiple planets around M dwarfs are more likely to be observed because they exist in more compact configurations, compared to those around FGK dwarfs. In other words, true FGK multis may be observed as singles in transit data because their planets are spaced wider apart. This would wash out the metallicity difference between singles and multis in the FGK case. Potentially related to this is an intriguing study by \citet{brewer2018b} that reports a significant difference in metallicity between singles and \emph{compact} multis around FGK dwarfs. This is the only study to uncover a metallicity difference between FGK singles and multis. Here, \emph{compact} is defined as systems with three or more planets orbiting within 1 au. It is possible that \citet{brewer2018b} detected a metallicity difference between multis defined as such and singles because they restricted themselves to FGK systems with compact planet configurations, which are less prone to such observational bias. In this scenario, FGK singles and multis also result from differential dynamics influenced by metallicity, but the multis are frequently observed as singles in transit data.

The underlying mechanism for the observable metallicity difference between singles and multis around M dwarfs, but not FGK dwarfs, is still unclear. However, there is ample evidence that the difference among M dwarf systems stems from planetary dynamics. We thus put forward the idea that M dwarf singles and multis are not drawn from the same underlying distribution, but rather are different populations with distinct dynamical histories.

%\section{Summary}
%In this paper, we used a sample of homogeneous chemical abundances to investigate metallicity differences between systems with one detected transiting planet (``singles") or multiple transiting planets (``multis"). We do not find any significant differences in metallicity between singles and multis around FGK dwarfs (in accordance with previous studies), but we do for systems around M dwarfs. Among M dwarf systems, multis are significantly more metal-poor than singles ($p$ = 0.0008). 

\section*{Acknowledgements}
%\begin{acknowledgments} % uncomment for line numbers
We thank Heather Knutson, Shreyas Vissapragada, and Claudia Aguilera-Gómez, as well as the CCA Exoplanet and Astro Data groups for supportive and productive conversations. %We also thank the anonymous referee for a helpful report. 

Funding for the Sloan Digital Sky Survey V has been provided by the Alfred P. Sloan Foundation, the Heising-Simons Foundation, the National Science Foundation, and the Participating Institutions. SDSS acknowledges support and resources from the Center for High-Performance Computing at the University of Utah. SDSS telescopes are located at Apache Point Observatory, funded by the Astrophysical Research Consortium and operated by New Mexico State University, and at Las Campanas Observatory, operated by the Carnegie Institution for Science. The SDSS web site is \url{www.sdss.org}.

SDSS is managed by the Astrophysical Research Consortium for the Participating Institutions of the SDSS Collaboration, including the Carnegie Institution for Science, Chilean National Time Allocation Committee (CNTAC) ratified researchers, Caltech, the Gotham Participation Group, Harvard University, Heidelberg University, The Flatiron Institute, The Johns Hopkins University, L'Ecole polytechnique f\'{e}d\'{e}rale de Lausanne (EPFL), Leibniz-Institut f\"{u}r Astrophysik Potsdam (AIP), Max-Planck-Institut f\"{u}r Astronomie (MPIA Heidelberg), Max-Planck-Institut f\"{u}r Extraterrestrische Physik (MPE), Nanjing University, National Astronomical Observatories of China (NAOC), New Mexico State University, The Ohio State University, Pennsylvania State University, Smithsonian Astrophysical Observatory, Space Telescope Science Institute (STScI), the Stellar Astrophysics Participation Group, Universidad Nacional Aut\'{o}noma de M\'{e}xico, University of Arizona, University of Colorado Boulder, University of Illinois at Urbana-Champaign, University of Toronto, University of Utah, University of Virginia, Yale University, and Yunnan University.

This work made use of data from the European Space Agency (ESA) mission Gaia (\url{https://www.cosmos.esa.int/gaia}), processed by the Gaia Data Processing and Analysis Consortium (DPAC, \url{https://www.cosmos.esa.int/web/gaia/dpac/consortium}). Funding for the DPAC has been provided by national institutions, in particular the institutions participating in the Gaia Multilateral Agreement.
\software{\texttt{numpy} \citep{numpy}, \texttt{matplotlib} \citep{matplotlib}, \texttt{pandas} \citep{pandas}, \texttt{scipy} \citep{scipy}, \texttt{scikit-learn} \citep{scikit-learn}, \texttt{astropy} \citep{astropy:2013, astropy:2018}, \texttt{JAX} \citep{bradbury2018}, \texttt{numpyro} \citep{phan2019,bingham2019}}
%\end{acknowledgments}

\bibliography{mybib}{}
\bibliographystyle{aasjournal}

\appendix
\setcounter{figure}{0}                       
\renewcommand\thefigure{A.\arabic{figure}}

\end{document}